\documentclass[12pt, letterpaper]{article} 
\usepackage[onehalfspacing]{setspace}
\usepackage [english]{babel}  
				
\usepackage[square]{natbib}	
\usepackage[margin=1in]{geometry}
\usepackage[mathscr]{euscript}
\usepackage{amssymb}
\usepackage{amsthm}
\usepackage{amsmath}
\usepackage{mathtools,eqparbox}
\usepackage{amsfonts}
\usepackage{csquotes}
\usepackage{epstopdf}
\usepackage{graphicx}
\usepackage{subcaption}
\usepackage{stackengine}
\usepackage{multirow,array}
\usepackage{booktabs}
\usepackage{bm}
\usepackage{comment}
\usepackage{float}

\newcommand{\indices}[2]{{
  \begin{array}{@{}r@{}}
    \scriptstyle #2~\smash{\eqmakebox[ind]{$\scriptstyle\rightarrow$}} \\[-\jot]  
    \scriptstyle #1~\smash{\eqmakebox[ind]{$\scriptstyle\downarrow$}}
  \end{array}}}

\usepackage{xpatch}
\usepackage{hyperref}
\hypersetup{colorlinks=true,citecolor=blue}

\title{A Statistical Equilibrium Approach to Adam Smith's Labor Theory of Value}
\author{Ellis Scharfenaker\footnote{Department of Economics, University of Utah, 260 Central Campus Drive, Gardner Commons, RM 4100, Salt Lake City, UT, 84112. Email: ellis.scharfenaker@economics.utah.edu. Tel: 801-581-7481}, Bruno Theodosio\footnote{Department of Economics, University of Tulsa.}, and Duncan K. Foley\footnote{Department of Economics, The New School for Social Research.}}

\date{\today}                                           

\begin{document}


\maketitle

\begin{abstract}
\singlespacing
Adam Smith’s inquiry into the emergence and stability of the self-organization of the division of labor in commodity exchange is considered using statistical equilibrium methods from statistical physics. We develop a statistical equilibrium model of the distribution of independent direct producers in a hub-and-spoke framework that predicts both the center of gravity of producers across lines of production as well as the endogenous fluctuations between lines of production that arise from Smith's concept of ``perfect liberty". The ergodic distribution of producers implies a long-run balancing of ``advantages to disadvantages" across lines of employment and gravitation of market prices around Smith’s natural prices.
\end{abstract}


\newpage
\section{Introduction}
    
Adam Smith's point of departure for understanding the rapid growth of labor productivity in 18th century England is an investigation of the self-organization of the division of labor in a system of specialized production and exchange. Smith introduces his theory of the division of labor in the first three chapters of the Wealth of Nations \citep{Smith1776} where it becomes the foundation for his theory of price in chapter six. Here, Smith introduces the ``early and rude" state of society in which a large number of independent direct producers who are free to decide what and how to produce might organize a division of labor capable of meeting the needs of social reproduction unlocking the social benefits of capitalism. Smith argues that the process that gives rise to the division of labor, increasing extent of the market, and increasing labor productivity, arises as a spontaneous outcome of decentralized decision making in the institutional context of free competition (Smith's ``perfect liberty").

There are two principal social coordination problems that Smith identifies in the first chapters of The Wealth of Nations. The first concerns the producer's decision problem of diversifying or specializing in production in order to meet their individual reproductive needs. The second concerns how specialized producers who must exchange their product as commodities with other specialized producers can organizing and sustain a division of labor capable of meeting the needs of social reproduction. Smith devotes some attention to the first problem in Books I and III, but assumes the economic conditions for specialized production and exchange tend to arise naturally from human's ``propensity to truck, barter, and exchange".\footnote{The problems of Neo-Smithinan historiography is well debated in the economic history literature for example, \citep{Brenner1987, Brenner2008, hiltonTransitionFeudalismCapitalism1982, Wood2002}.}  Smith's main interest is in understanding the underlying laws and tendencies of systems where specialized producers must exchange products as commodities against money in order to meet their individual needs of reproduction. In this setting, the fluctuations of payoff-maximizing producers across lines of production leads Smith to a theory of value and prices based on the labor expended in production. 

To illustrate his theory of value and price, Smith presents an abstract thought experiment in which specialized producers who own or create their own means of production and are free to move from one line of production to another will tend to ``balance the advantages to disadvantages" across all lines of production. When the primary disadvantage of production is the average labor effort and primary advantage the money income the production affords the producer, the movement of producers will tend to equalize the ratio of income to labor effort across lines of production. Because producers do not coordinate these decisions and there is no social planner organizing the division of labor, there will be a considerable element of chance in finding any particular producer in any particular line of production at any point in time. Smith recognizes that the problem of achieving the social division of labor is irreducibly statistical and primarily concerns the stability and (long-run) equilibrium conditions of the distribution of  producers \citep{ScharfYang2020}. Smith is clear in his statistical description of commodity production that the endogenous movement of producers among different lines of production results in meeting the needs of social reproduction only \textit{on average}. 

Smith's theory of value and prices in its simplest form concerns two feedbacks that we can model as conditional probability distributions. The first is that independent producers who take the payoff in particular sectors of production as determined by forces beyond their control will move from sectors with relatively low payoffs to those sectors with relatively higher payoffs, where payoffs are understood as their income relative to labor effort (``advantages" relative to ``disadvantages"). 

\begin{quote}
    The whole of the advantages and disadvantages of the different employments of labour and stock must, in the same neighbourhood, be either perfectly equal or continually tending to equality. If in the same neighbourhood, there was any employment evidently either more or less advantageous than the rest, so many people would crowd into it in the one case, and so many would desert it in the other, that its advantages would soon return to the level of other employments.\citep[pp.142]{Smith1776}
\end{quote}

The second feedback mechanism is that the movement of producers into (out of) a sector tends to lower (raise) the payoff through the dual movement of prices and incomes. 

\begin{quote}
    The occasional and temporary fluctuations in the market price of any commodity fall chiefly upon those parts of its price which resolve themselves into wages and profit... Such fluctuations affect both the value and the rate either of wages or of profit, according as the market happens to be either overstocked or understocked with commodities or with labour.\citep[pp.88]{Smith1776}
\end{quote}

Both of these factors are essential to the process of convergence of market prices to natural prices. If individual producers didn't pay any attention to the expected payoff in deciding where to produce, there would be no tendency toward the convergence of market prices to natural prices. If the movement of producers into or out of a sector had no impact on the prices and incomes in the sector through competition, even if producers do seek the highest rates of return, their actions would also not lead to a tendential convergence of prices. These considerations indicate that we need to think of Smith's theory of value in terms of an equilibrium joint frequency distribution over the the distribution of producers among sectors of production and the actions of producers that constitute their movement among sectors. We show in the simplest setting that these two parts of Smith's thought experiment imply a Markov process describing the stochastic movement of producers with an ergodic distribution that on average ``balances the advantages and disadvantages" of production and implies market prices will gravitate around natural prices proportional to the labor embodied in commodities.

\section{Hub-and-Spoke Model}

\begin{figure}[ht!]
\begin{center}
\includegraphics[scale = 0.5]{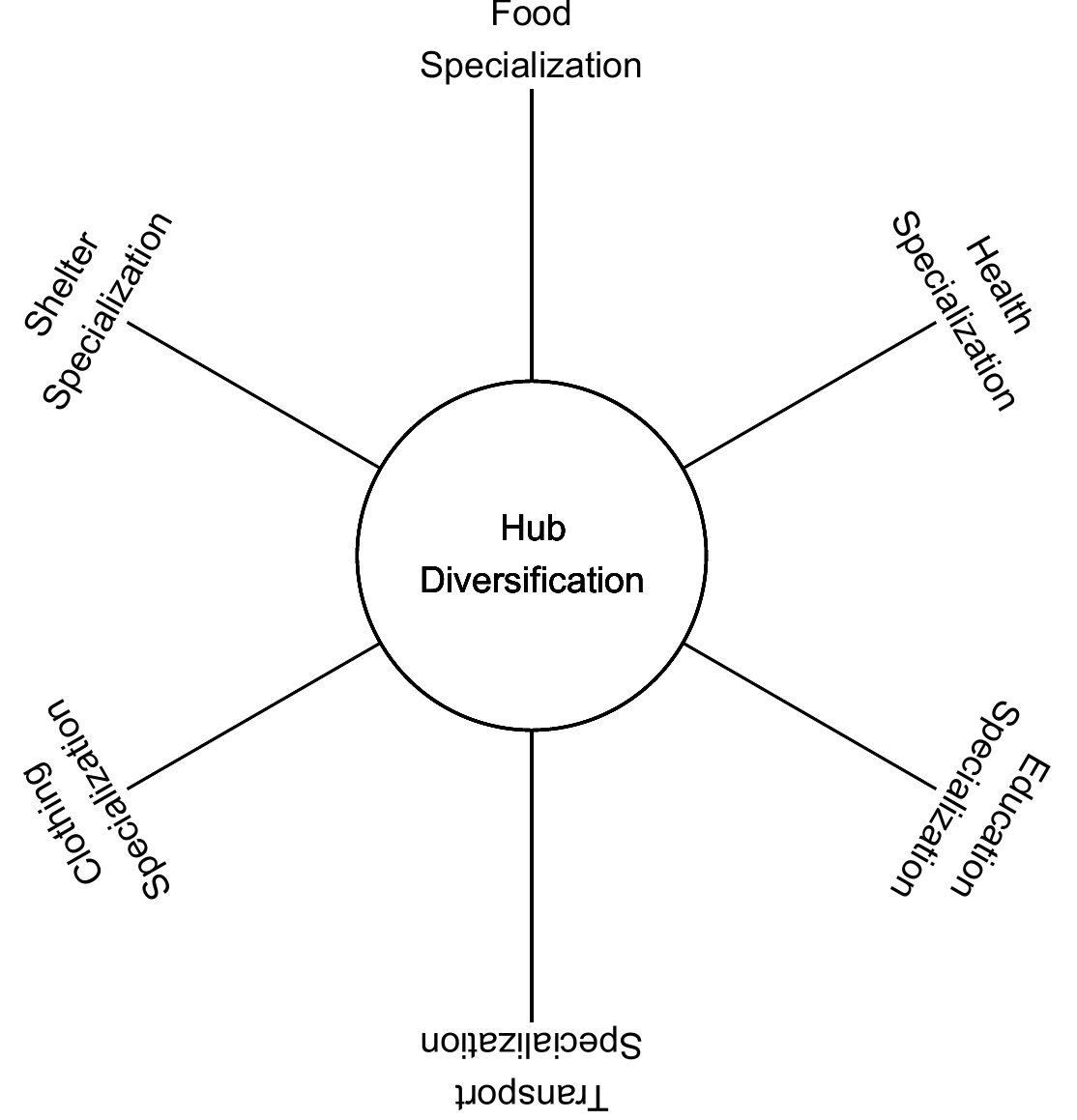}
\caption{The hub-and-spoke representation of the independent producers decision to diversify and produce all needs for individual reproduction at the hub or specialize in the production of a good, or range of goods in a spoke, that only partially meets their individual needs of  reproduction. Figure reproduced from \citep{foleySocialCoordinationProblems2020}.}
\label{fig:HubSpoke}
\end{center}
\end{figure}

The problem of organizing the social division of labor can be represented abstractly through a hub-and-spoke model \citep{foleySocialCoordinationProblems2020} illustrated in Figure~\ref{fig:HubSpoke}. This model exemplifies the two-fold problem of producers choosing to diversify or specialize in production and how to distribute themselves among the spokes in order to meet the needs of social reproduction. In the hub-and-spoke model each producer faces the choice of diversifying production at the hub, and producing all of their needs on their own, or specializing in the production of a narrow range of products in one of the spokes. A diversified producer will be largely self-sufficient and the exchange of products with other producers will be of secondary importance to the allocation of labor within the producing unit (think Robinson Crusoe). A producer who specializes at a spoke must produce a surplus of one particular product and exchange it for a variety of other products in order to meet their needs of individual reproduction. Specialized production requires the existence of other differently specialized producers to sustain individual reproduction through the exchange of products. Because specialized producers produce with the intention of exchanging the surplus of products take the form of exchange values. Smith's theory of value is a theory of the fluctuations and long-term distribution of specialized producers across the spokes of production.

In the hub-and-spoke model  there are $N>>1$ identical independent producers and $K$ spokes, each representing a socially necessary line of production. The system is described as a vector $\{n_1,n_2,\cdots,n_K\}$, where $n_k$ is the number of producers in spoke $k$, and $\sum_{k=1}^{K} n_k=N$ is the total number of producers that define the degrees of freedom of the system.  Figure~\ref{fig:HubSpoke} illustrates the hub-and-spoke model for $6$ different lines of production. 

The hub-and-spoke model is a useful metaphor for Smith's social coordination problem of achieving the social division of labor through specialized production.\footnote{``As it is by treaty, by barter, and by purchase that we obtain from one another the greater part of those mutual good offices which we stand in need of, so it is this same trucking disposition which originally gives occasion to the division of labour. In a tribe of hunters or shepherds a particular person makes bows and arrows, for example, with more readiness and dexterity than any other. He frequently exchanges them for cattle or for venison with his companions; and he finds at last that he can in this manner get more cattle and venison than if he himself went to the field to catch them. From a regard to his own interest, therefore, the making of bows and arrows grows to be his chief business, and he becomes a sort of armourer. Another excels in making the frames and covers of their little huts or movable houses. He is accustomed to be of use in this way to his neighbours, who reward him in the same manner with cattle and with venison, till at last he finds it his." (\citep{Smith1776}, pp.31)} 

\subsection{Specialized Production}
In the hub-and-spoke model producers' decisions to locate themselves at any particular spoke depends on the shape of the feasible frontier, the relative prices of the goods, and on the social and institutional context in which the producers find themselves. We assume that specialized production and exchange is a Cournot-Nash equilibrium and that independent producers face no physical or institutional barriers to exchanging with one another. The system is assumed to have strong strategic complementarity in the decision of producers choice to specialize and exchange rather than diversify in production at the hub. The question is, what determines the conditions of exchange and how can an economy of specialized producers sustain a social division of labor that meets the needs of social reproduction? 

In the simplest setting agents produce and consume two perishable goods, sugar $(s)$ and corn $(c)$ and need both in a fixed proportion to survive. The assumption of perishability precludes the possibility of the accumulation of goods over time. Producers' payoffs are described by the Leontief function in the form $\min[c,s]$. Because producers specialize in the production of one good they must exchange their surplus for other goods at an exchange ratio. We can choose the units in which producers require each good as being a proportional $1:1$. The Leontief expression of proportional needs reflects the classical assumption of historically determined exogenous social needs of reproduction for a given period.  
 
\begin{figure}[H]
\begin{center}
\includegraphics[]{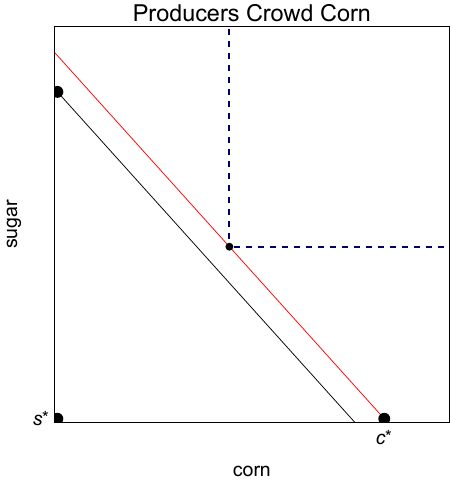} \includegraphics[]{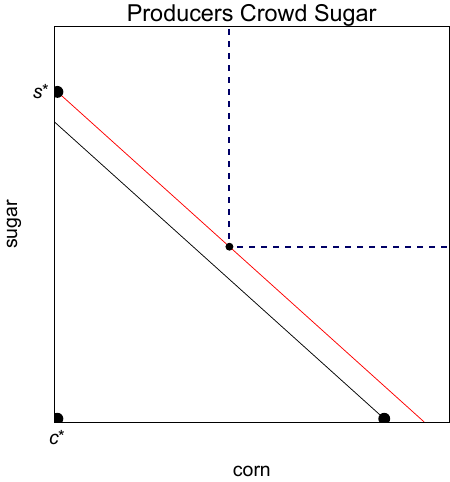}
\caption{Two price systems, one in which it is advantageous to produce corn (left) and one in which it is advantageous to produce sugar (right). When payoffs are higher for corn production there is a high probability that sugar producers will move to corn production. The opposite will happen when there is an advantage to sugar production.}
\label{fig:ppf}
\end{center}
\end{figure}

Smith argues in Ch.6 that an economy of specialized producers who must exchange in order to meet their individual needs of reproduction will almost certainly adopt a common commodity as money due to the attendant network externalities \citep{foleySocialCoordinationProblems2020}. If we take sugar as the money commodity, the price of corn in terms of sugar is $p=\frac{p_c}{p_s}$. This ratio defines a linear price-exchange surface on which a specialized producer will be able to move from a boundary solution to an interior optimum. The price-exchange line through the feasibility frontier (endowment) defines the individual producer's budget constraint and the negative of the slope of this line is the sugar price of corn. If labor is indivisible and each producer must decide whether to allocate their labor time to corn or sugar production, the production possibility frontier will consist of three points, two boundary points representing production at each spoke and an origin point representing zero production.\footnote{For simplicity we assume that diversification is not a feasible strategy in an economy of specialized producers.} A price system will define payoffs at each spoke as in Figure~\ref{fig:ppf}.

If prices are such that there is a relative advantage for producing corn the market will not clear and producers who specialize in sugar will begin to exit sugar and crowd corn production. As producers crowd corn production the price of corn and payoff for producing corn will fall. As producers move away from sugar production the price of sugar and payoff for producing sugar will increase. If, for example, the new prices are such that the expected payoff from sugar production becomes higher than corn the typical producer will then exit corn and enter sugar production. These two cases are illustrated in the left and right panels of Figure~\ref{fig:ppf}. Smith acknowledges that this migration of labor is an interminable process that does not have any tendency to settle down to a steady state fixed-point equilibrium in which producers are stationary. 

The movement of producers, however, does result in the formation of a center of gravity that Smith identifies with natural prices in simple commodity production, represented by the line connecting the two specialization corner solutions. At natural prices commodities will exchange at their labor values and $p \propto \frac{\lambda_c}{\lambda_s}$.

\begin{figure}[H]
\begin{center}
\includegraphics[]{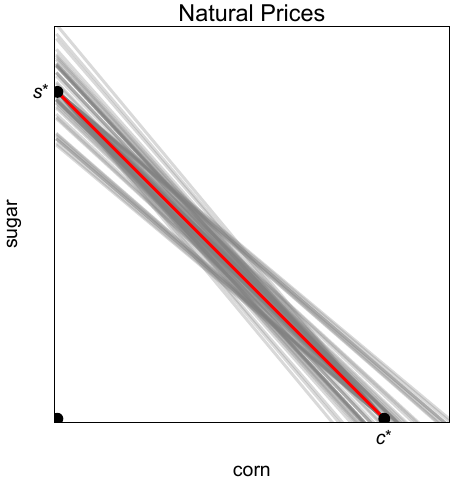}
\caption{The concept of natural prices is illustrated by the red line connecting the two spokes of the feasibility frontier. The gray lines illustrate fluctuations of market prices around the natural price. At natural prices, the payoffs are equalized and the market clears.}
\label{fig:Natprices}
\end{center}
\end{figure}

At any one moment we should expect that market prices will deviate from natural prices, but from the continuous migration of labor responding to the changes in relative advantage of employment, a center of gravity emerges around competitive natural prices. As Smith explains, ``The natural price is the central price to which the prices of commodities are continually gravitating. Different accidents may sometimes keep them suspended a good deal above it, and sometimes force them down even somewhat below it. But whatever may be the obstacles which hinder them from settling in this center of repose and continuance, they are constantly tending towards it."\citep[pp.87]{Smith1776} Figure~\ref{fig:Natprices} illustrates the gravitational process of market prices (gray lines) around the natural price that is proportional to the labor effort required in each line of production (red line connecting the spokes).

Smith emphasizes that even if the incentives are balanced across all spokes an unbalanced allocation of producers will still persist due to the decentralized decision making of independent producers. Smith is clear in his argument that the balanced staffing over spokes will occur only \textit{on average} over many cycles of production. 

\begin{quote}
``It is only the average produce of the one species of industry which can be suited in any respect to the effectual demand; and as its actual produce is frequently much greater and frequently much less than its average produce, the quantity of the commodities brought to market will sometimes exceed a good deal, and sometimes fall short a good deal, of the effectual demand. Even though that demand therefore should continue always the same, their market price will be liable to great fluctuations, will sometimes fall a good deal below, and sometimes rise a good deal above their natural price. In the other species of industry, the produce of equal quantities of labour being always the same, or very nearly the same, it can be more exactly suited to the effectual demand. While that demand continues the same, therefore, the market price of the commodities is likely to do so too, and to be either altogether, or as nearly as can be judged of, the same with the natural price."\citep[pp.87-88]{Smith1776}
\end{quote}

Because the predominant disadvantage of an employment is the labor time required to conduct it, the process of equalization of incomes implies natural competitive prices of commodities will be proportional to their ``embodied labor time". Smith's labor theory of value is both an account of the formation of natural prices as well as an account of the distribution of social labor. 

In the next section we develop a quantitative picture of Smithian equilibrium by using methods from statistical physics to model the spontaneous formation of the division of labor in terms of a statistical equilibrium distribution of producers.

\section{Division of labor with two perishable goods}

We consider a large number of producers, $N>>1$, categorized into two distinct groups based on their production: those producing corn, represented as $N_c$, and those producing sugar, represented as $N_s$. During a given production period, it is assumed that the total quantity of corn ($X$) and sugar ($Y$) produced is equivalent to the number of producers in each respective category. Consequently, the per-producer production of corn ($n_c$) and sugar ($n_s$) can be calculated as $n_c = \frac{X}{N}$ and $n_s = \frac{Y}{N}=1-n_c$,  respectively. We follow Smith and assume that if any intermediate goods are necessary for production they will emerge in terms of the final good and do not assume the social relations of capital. We also assume the social demand is equal for sugar and corn.

Following \citep{bowlesMicroeconomicsCompetitionConflict2022} we refer to markets that do not clear as characterized by a ``long-side" and a ``short-side". On the ``long-side", either supply or demand, has a greater number of desired transactions at a given price. Not all producers on this side will be able to complete their desired transactions, leading to some being quantity constrained. On the other hand, the ``short-side", which can also be either supply or demand, has fewer desired transactions at that given price. Producers on the ``short side" are able to complete all their desired transactions and retain a ``short-side" power over the ``long side".

We assume that after a production period the prices for a good produced on the ``long side" of the market will become so small that the producers on the short side will effectively receive the entire surplus from market exchange. 

In the hub-and-spoke model an excess supply of corn implies $n_c >\frac{1}{2}$ in which case corn will be on the ``long side" and sugar on the ``short side". Symmetrically, if there is excess supply of sugar and $n_c<\frac{1}{2}$, sugar will be on the ``long side" and corn the ``short side". 

If at the end of a production period there is excess supply of sugar ($n_c <\frac{1}{2}$) the price of sugar will become so small that a sugar producer will get zero corn and zero sugar for payoff $\min[0,0] = 0$. Corn producers on the ``short side" will get the $\frac{X}{N_c} = 1$ unit of corn they produce as well as some fraction of the sugar produced. If we assume that each corn producer gets an equal average amount of sugar then $\frac{Y}{N_c} = \frac{n_s}{1-n_s} = \frac{1-n_c}{n_c}$. In this case, a typical corn producer's payoff is $\min[1,\frac{1-n_c}{n_c}] = 1$. Equivalently, we can say that the ``typical" producer will have a payoff $\min[0,0] = 0$ with probability $1-n_c$ and a payoff $\min[1,\frac{1-n_c}{n_c}] = 1$ with probability $n_c$. Schematically, the typical producer faces the following payoffs when corn is on the ``short side" of the market:

\begin{equation}
\text{$n_c < \frac{1}{2}$}
\begin{cases}
    \text{With Prob.} = n_c 
    \begin{cases}
        \text{Corn: } \frac{X}{N_c}=1
        \\
        \text{Sugar: } \frac{Y}{N_c} = \frac{n_s}{1-n_s} = \frac{1-n_c}{n_c}
    \end{cases}
&\rightarrow \min\left[1, \frac{1-n_c}{n_c}\right] =1
    \\
    \\
    \text{With Prob.} =1-n_c
    \begin{cases}
        \text{Corn: } 0
        \\
        \text{Sugar: } 0
    \end{cases}
    &\rightarrow \min[0,0] =0
\end{cases}
\end{equation}

When corn is on the ``short side" and sugar is on the ``long side", there will be a total excess supply of sugar equal to $Y-N_c>0$ and per-producer excess supply of $1-2n_c > 0$ that will be disposed of at the end of the period. No accumulation is possible when both goods are perishable.

By symmetry of the goods, when sugar is on the ``short-side" and $n_c >\frac{1}{2}$, assuming equal distribution of corn among sugar producers the typical producer will face the following payoffs:

\begin{equation}
\text{$n_c > \frac{1}{2}$}
\begin{cases}
    \text{With Prob.} =n_c
    \begin{cases}
        \text{Corn: } 0
        \\
        \text{Sugar: } 0
    \end{cases}
    &\rightarrow \min[0,0] =0
    \\
    \\
     \text{With Prob.} = 1-n_c 
    \begin{cases}
        \text{Corn: } \frac{X}{N_s}=\frac{n_c}{1-n_c} 
        \\
        \text{Sugar: } \frac{Y}{N_s} = 1
    \end{cases}
&\rightarrow \min\left[\frac{n_c}{1-n_c},1\right] =1
\end{cases}
\end{equation}

In Smith's system, direct commodity producers respond to payoff differences by moving from disadvantageous lines of production to advantageous lines of production. The decentralized nature of decision making, however, implies the typical producer will move from one line of production to another probabilistically. A simple and parsimonious way of modeling the partial randomization of strategies is by constraining the typical producer's mixed strategy with a minimum informational entropy \citep{foleyInformationTheoryBehavior2020, scharfenakerImplicationsQuantalResponse2020}. In the hub-and-spoke model producers must choose to produce at one of the $K$ spokes representing the production of specific commodities but has limited information on where other producers are located. The limited information that arises from decentralized production is a key constraint in Smith's theory. To assume that producers face no information processing constraints on short-run production decisions is consistent with two equilibrium solutions: ceaseless fluctuations of all producers between lines of production, since infinitesimally small differences in payoffs will induce an instantaneous avalanche of producers into the more profitable sector, or an unstable interior solution of an equal distribution of producers at each spoke, which might be maintained by a central authority. Smith, however, is interested in how the spontaneous organization of the social division of labor arises in a decentralized economy of independent producers. Answering this question requires a probabilistic framework for modeling the division of labor in a dynamic context.

We can represent the producers' decision problem as choosing an action $\{a_1,\cdots,a_K\}$ representing the specialization of production at a spoke knowing the payoff $u[a_k]$ associated with each action. We can represent the typical producer's mixed strategy in terms of a frequency distribution $\{f_1,\cdots,f_K\}$, $\sum_k f_k=1$, and expected payoff $\sum_k f_k u[a_k]$. Maximizing the expected payoff subject to only the normalization of frequencies implies that the producer will choose the line of production with the highest payoff with certainty. In this situation the social division of labor is unattainable in any period of production since all producers will always crowd into a single line of production. If we constrain producers' mixed strategy with a minimum informational entropy the maximization problem for the typical producer can be expressed as:

\begin{equation} 
\begin{split}
\max_{\left\{ f_k \geq 0 \right\}} \sum_k f_k u[a_k] \quad \\
\text{subject to} \quad \sum_k f_k = 1 \\
- \sum_k f_k \log[f_k] \geq H_{\min}
\end{split}
\end{equation}

The Lagrangian associated with this maximization problem, with the two Lagrangian multipliers $T$ and $\mu$ is:
\begin{equation} 
\begin{split}
\mathcal{L}[f_k;\mu,\lambda] = -\sum_k f_k u[a_k] - \mu \left( \sum_k f_k - 1 \right) \\
+ T \left( -\sum_k f_k \log[f_k] - H_{\min} \right)
\end{split}
\end{equation}

The Lagrange multiplier $\mu$ ensures the normalization of the frequencies over actions. We refer to $T$ as the ``predictable behavior scale" (PBS) since it measures the scale of fluctuations of individual behavior resulting from the sensitivity of agents to differences in payoffs.\footnote{The PBS has a parallel interpretation as the temperature in physical systems.} A lower $T$ makes agents more sensitive to differences in payoffs as the constraint is less binding. The solution to the constrained maximization problem is the Gibbs (SoftMax) distribution over actions. 
\begin{equation}
    f[a_k]=\frac{e^{\frac{u[a_k]}{T}}}{\sum_k e^{\frac{u[a_k]}{T}}}
\end{equation}
When there are just two actions, such as $a_1=\text{corn}$ and $a_2=\text{sugar}$, the solution reduces to the logit quantal response function:

\begin{equation}
    f[a_c]=\frac{e^\frac{u[a_c]}{T}}{e^\frac{u[a_c]}{T}+e^\frac{u[a_s]}{T}} = \frac{1}{1+e^\frac{u[a_s]-u[a_c]}{T}}
        \end{equation}
\begin{equation}
    f[a_s]=1-f[a_c]= \frac{1}{1+e^{-\frac{u[a_s]-u[a_c]}{T}}}
        \end{equation}

The Gibbs distribution implies producers will choose each line of production with a positive frequency. For two actions, such as corn and sugar, the relative logs odds of choosing an action is just the difference in payoffs scaled by $T$:

\begin{equation}
    \log \left[\frac{f_s}{f_c}\right]=\frac{u[a_s]-u[a_c]}{T}
\end{equation}

If corn producers' expected payoff for producing in period $t+1$ is just their payoff from producing at time $t$, they will move from one line of production to another with probability:
\begin{equation}
    f\left[n_c[t+1]|n_c[t]\right] =
\begin{cases}
    \frac{1}{1+e^\frac{\min[0,0]-\min \left[1, \frac{1-n_c[t]}{n_c(t)}\right]}{T}} = \frac{1}{1+e^{-\frac{1}{T}}}\ \ \ \ \text{if} \ n_c[t]<\frac{1}{2}
    \\
    \frac{1}{1+e^{-\frac{\min\left[1, \frac{1-n_c[t]}{n_c[t]}\right]-\min[0,0]}{T}}} = \frac{1}{1+e^\frac{1}{T}}\ \ \ \ \text{if} \ n_c[t]>\frac{1}{2}
\end{cases}
\end{equation}

\begin{figure}[ht!]
\begin{center}
\includegraphics[]{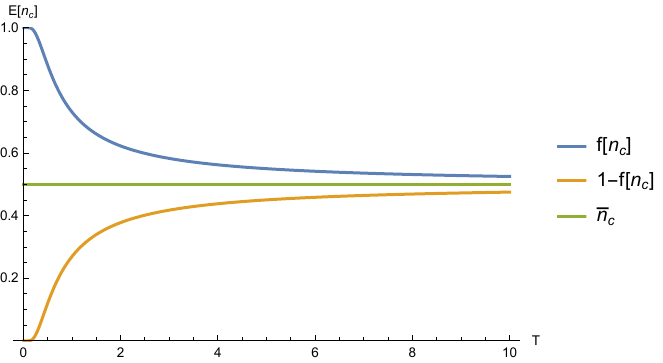}
\caption{The predicted frequency of corn and sugar producers as a function of the predictable behavioral scale $T$. When $T\rightarrow0$ all producers crowd into a single spoke and cannot sustain a division of labor. When $T\rightarrow \infty$ payoff differences will not induce movement of producers and producers will essentially behave randomly. $\bar{n}_c$ is the mean number of corn producers.}
\label{fig:Tempplot}
\end{center}
\end{figure}

We can visualize the expected number of producers in each line of production as a function of the ``predictable behavior scale" $T$ in Figure~\ref{fig:Tempplot}. 

Because $T$ represents the informational processing constraint of individual producers, when $T \rightarrow 0$ producers are infinitely sensitive to payoff differences and will move deterministically, instantaneously crowding into the line of production with the highest payoff. As $T \rightarrow \infty$ producers decisions become independent of payoff differences and will be equally likely to choose either line of production. In this case producers behave ``randomly" in their production decisions choosing sugar or corn by the flip of a coin. The purposive decentralized nature of decision making in unplanned competitive market economies is neither likely to approximate the deterministic ``degenerate" case in which producers switch lines of producing for infinitesimal differences in payoffs nor the completely random case in which producers are completely unresponsive to payoffs. We should expect, as Smith did, that producers operate between these two extremes, where payoffs incentivize producers to move, but the decentralized market interactions make this movement probabilistic.

\subsection{Dynamics and statistical equilibrium}
The stochastic quantal response of the typical producer induces a Markov chain on the state space of profiles of agent behavior. If there are $N$ producers each with the same PBS, $T$, the state of the system, that describes the distribution of producers is described by the number of producers choosing to produce corn, $N_c=0,1,...,N$ and the average frequency of taking the choosing corn production will be $\frac{N_c}{N}=n_c$. The frequency with which each producer will choose to produce corn is $f[n_c]=\frac{1}{1+e^{-\frac{1}{T}}}$ if $n_c<1/2$ and $1-f[n_c]=\frac{1}{1+e^{\frac{1}{T}}}$ if $n_c>1/2$ and the transition probabilities from one state to another for each agent takes the Binomial form:

\begin{equation}
\mathcal{B}_{N,N_c}[f[n_c]]={N \choose {N_c}}f[n_c]^{N_c}{(1-f[n_c])}^{N-N_c}
\end{equation}

A Markov transition matrix that allows for zero corn production can be constructed by assuming there are $N+1$ producers each described by a Binomial transition probability\footnote{The Binomial distribution is defined for $\{N_c=0,1,2,\cdots,N\}$ which includes the state of no corn production $N_c=0$; therefore, in order to construct a square transition matrix it is necessary to consider an economy of $N+1$ producers. Assuming $N$ producers does not change the qualitative features of the model, it only implies that at least one producer will choose corn production.}:

\begin{table}[ht!]
\[
  \begin{array}{c|ccccc}
    \indices{n}{N_c}
    & 0 & 1 & 2 & \cdots & N \\
    \hline
    0 & \mathcal{B}_{N,0}[f[n_c]] & \mathcal{B}_{N,1}[f[n_c]] & \mathcal{B}_{N,2}[f[n_c]] & \cdots & \mathcal{B}_{N,N}[f[n_c]] \\
    1 & \mathcal{B}_{N,0}[f[n_c]] & \mathcal{B}_{N,1}[f[n_c]] & \mathcal{B}_{N,2}[f[n_c]] & \cdots & \mathcal{B}_{N,N}[f[n_c]] \\
    2 & \mathcal{B}_{N,0}[f[n_c]] & \mathcal{B}_{N,1}[f[n_c]] & \mathcal{B}_{N,2}[f[n_c]] & \cdots &\mathcal{B}_{N,N}[f[n_c]]\\
    \vdots & \vdots & \vdots & \vdots & \ddots & \vdots\\
    \frac{N+1}{2} & \mathcal{B}_{N,0}[1-f[n_c]] & \mathcal{B}_{N,1}[1-f[n_c]] & \mathcal{B}_{N,2}[1-f[n_c]] & \cdots & \mathcal{B}_{N,N}[1-f[n_c]]\\
    \vdots & \mathcal{B}_{N,0}[1-f[n_c]] & \mathcal{B}_{N,1}[1-f[n_c]] & \mathcal{B}_{N,2}[1-f[n_c]] & \cdots & \mathcal{B}_{N,N}[1-f[n_c]]\\
    N & \mathcal{B}_{N,0}[1-f[n_c]] & \mathcal{B}_{N,1}[1-f[n_c]] & \mathcal{B}_{N,2}[1-f[n_c]] & \cdots & \mathcal{B}_{N,N}[1-f[n_c]]\\
  \end{array}
\]
\caption{The transition kernel for $N$ producers is a right stochastic matrix characterized by the Binomial distribution where the frequency of choosing corn production is described by the logit quantal response function. The parameter $T$ controls the sensitivity of producers' decisions to payoff differences between the two spokes.}
\end{table}

Because the Markov transition matrix is an $(N+1)\times (N+1)$ irreducible, non-negative stochastic matrix, the Perron-Frobenius theorem tells us that there exists a unique real eigenvalue (equal to one) with a corresponding positive eigenvector, that when normalized is the ergodic distribution of the Markov chain. In this case, the ergodic distribution is the stationary distribution of producers as the number of production periods becomes large.

\begin{figure}[H]
\begin{center}
\includegraphics[scale=1.1]{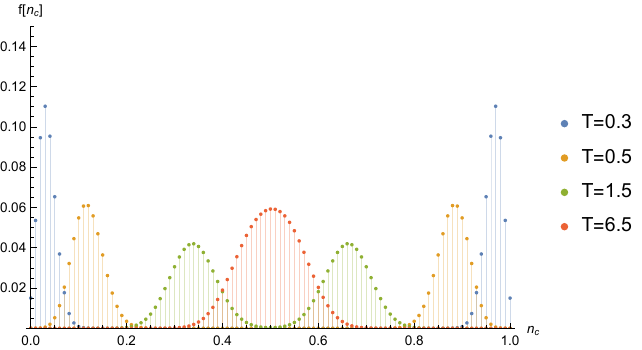}
\caption{The ergodic distribution of producers for different predictable behavior scales. When $T$ is small producers are more sensitive to differences in payoffs and will fluctuate quickly from one line of production to the other. As $T$ approaches the zero limit the ergodic distribution predicts that all producers will either be crowded into corn or sugar production. When $T$ is large producers are insensitive to differences in payoffs and will have an equal or uniform probability of producing corn or sugar.}
\label{fig:ergodic}
\end{center}
\end{figure}

Figure~\ref{fig:ergodic} represents the ergodic distribution for $N=100$ producers for different values of $T$. When $T$ is small the ergodic distribution is bi-modal and concentrates on extreme distributions in which most producers choose either to produce corn or steel. When $T$ is small there is also a positive frequency of transition between the extreme configurations of the system. In the limit as $T\rightarrow 0$ the ergodic distribution concentrates entirely on the extreme outcomes and all producers will entirely crowd corn or sugar depending on the relative advantage. In this deterministic case there is always a degenerate distribution of producers in one spoke that cannot sustain a social division of labor. For positive values of $T$ producers will fluctuate between the two spokes. For high values of $T$ the ergodic distribution is centered on an interior unimodal equilibrium in which payoff differences between the spokes will have no impact on producers decisions. When producers do respond to differences in payoffs and face informational constraints due to the decentralized nature of production $T$ will correspond to an intermediate case in which producers ceaselessly fluctuate between lines of production giving rise to a center of gravity at which  the expected number of producers $E[n_c]$ corresponds to Smith's balanced ``advantages and disadvantages" of employment. On average the labor theory of value and the spontaneous organization of production through specialization is a stochastically stable equilibrium \citep{Young1993}.

\subsection{Inequality}
\begin{figure}[H]
\begin{center}
\includegraphics[scale=1]{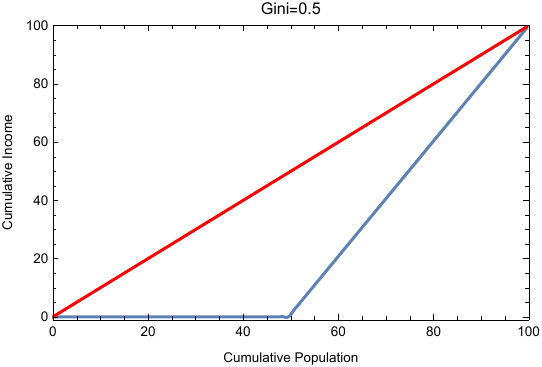}
\caption{The Lorenz curve for the perishable goods economy at the ergodic distribution of producers. The Gini coefficient is equal to $0.5$.}
\label{fig:inequality}
\end{center}
\end{figure}

While the perishable goods economy is a simple highly stylized model the division of labor and the spontaneous emergence of production through specialized production, it does demonstrate that production and exchange will endogenously generate inequality. Figure~\ref{fig:inequality} shows the Gini index and a Lorenz curve for this model for $N=100$ producers, showing the cumulative income as a function of the cumulative population of corn producers for the long-run average distribution of producers.

The blue line is the Lorenz curve corresponding to the  cumulative percentage of the population and the cumulative percentage of goods owned by that portion of the population. The red line is the line of perfect equality, which represents the socially coordinated outcome in which producers specialize and divide the net product equally. The Gini coefficient, ratio of the area between the Lorenz curve and $45-$degree line to the total area under the $45-$degree line is $0.5$ for decentralized specialized production.

\section{Discussion}

Smith's ``gravitational" equilibrium theory of natural prices has been predominantly articulated within a ``long-period" framework that acknowledges the endogenous fluctuations of market prices, but because of ``how intrinsically complex the issue of gravitation is"\citep{Kurz1995} abstracts away from these statistical complexities. One unfortunate effect of the success of the ``long-period" approach was that it turned attention away from the development of a coherent statistical methodology in classical political economy. We emphasize that statistical equilibrium is not an approximation to idealized systems operating at zero entropy. The conclusions one might draw from the ``stylized" economy are often not supported in statistical equilibrium \citep{ScharfenakerFoley2024}. In this paper we demonstrate how a division of labor can spontaneously self-organize and be sustained over time due to the fluctuations of producers across lines of production. Producers behave purposively in their decisions seeking out the highest rate of remuneration. The decentralized nature of market interactions induces a minimum entropy on producers' action set leading to a non-degenerate equilibrium distribution of producers across all lines of production.

Non-degenerate prices and incomes in equilibrium also tends to be understood in much of the existing literature in terms of a system in disequilibrium. The movement of agents in statistical equilibrium, however, is  conceptually distinct from the study of disequilibrium dynamics (such as those studied in \citet{DumenilLevy1991}). A statistical equilibrium model substitutes a probabilistic description of the system, in terms of the configurations of the system, for a detailed dynamic prediction of the movement of each individual part. A system in statistical equilibrium is defined by a frequency distribution over all states of the system. In the hub-and-spoke framework this implies an equilibrium frequency distribution of producers consistent with Smith's theory of value.

As \citep[pp.20]{Kurz1995} acknowledge, ``A proper answer to it [how to model gravitational equilibrium] would seem to contain, of necessity, an answer to many economic questions which are as yet unresolved." We believe that the statistical equilibrium approach to Smith's gravitational equilibrium proves a step in the right direction to answering this question. Articulating a theory of the statistical effects of market mediated production and exchange in systems with many agents interacting in complex ways is fundamental to revealing and understanding the statistical regularities in economic data. 

While we have focused our statistical analysis of the hub-and-spoke model on the social division of labor in an economy of direct producers (simple commodity production) the same ideas can be extended to a capitalist production where the means of production (and subsistence) are owned by a class other than direct producers. In this case, as long as labor is still free to move among different employments, even though it must become employed of the owners of the means of production working for a wage, the free mobility of labor will still enforce a balance between the advantages and disadvantages of income. If the employers who appropriate a profit also move their capital among different employments seeking the highest rate of profit the centers of gravity will tend to equalize rates of profit for capital and the ratio of labor effort to money wages for labor. 

A further development of the model could also include an economy with a durable goods, such as steel. In this case steel can be accumulated over time and would generate dynamics of inequality distinct from the perishable goods economy.

\section{Conclusion}

Classical Political Economy recognized capitalism as a complex social system. The astronomical degrees of freedom, complex interdependencies, and numerous feedbacks make modeling complex systems a formidable task. Because of the complexity of the system classical political economists such as Adam Smith argued that any conclusions drawn about capitalism must rest on robust, pervasive, self-reinforcing (statistical) tendencies. The emergence of natural prices from the free mobility of independent producers is a simple yet powerful illustration of Smith's methodological approach. Smith's logic concerning the process of the spontaneous formation of the social division of labor and its implications for the theory of value is inherently statistical. Centers of gravity in prices emerge through the endogenous fluctuations of individual producers between different lines of production. Free competition and the decentralized nature of production decisions implies the movement of producers among different lines of production is stochastic. We address this irreducible element of randomness in Smith's theory by developing a statistical equilibrium hub-and-spoke model in which entropy constrained direct specialized producers balance the ``advantages and disadvantages" of employment through their stochastic movement between spokes. The resulting ergodic distribution of producers supports Smith's theory of gravitational equilibrium and the labor theory of value.

\section*{Acknowledgement}
We would like to acknowledge helpful comments from Heinz Kurz and Bertram Schefold, as well as helpful discussions with participants of the 2023 International Conference on Economic Theory and Policy at Meiji University, Japan. 
\clearpage 
\bibliography{/Users/u0687358/Dropbox/Master.bib}

\end{document}